\documentclass[%
 aip,
 amsmath,amssymb,
 reprint,%
]{revtex4-1}

\usepackage{graphicx}
\usepackage{dcolumn}
\usepackage{bm}

\usepackage[utf8]{inputenc}
\usepackage[T1]{fontenc}
\usepackage{mathptmx}
\usepackage{etoolbox}
\usepackage{upgreek}
\usepackage{float}

\newcommand{\mb}[1]{\mathbf{#1}}
\newcommand{\br}[1]{\left [ #1 \right ]}

\newcommand{\uN}{$\upmu$N~}

\makeatletter
\def\@email#1#2{%
 \endgroup
 \patchcmd{\titleblock@produce}
  {\frontmatter@RRAPformat}
  {\frontmatter@RRAPformat{\produce@RRAP{*#1\href{mailto:#2}{#2}}}\frontmatter@RRAPformat}
  {}{}
}%
\makeatother
\begin{document}

\preprint{AIP/123-QED}

\title[Direct measurement of forces in air-based acoustic levitation systems]{Direct measurement of forces in air-based acoustic levitation systems}

\author{Nina M.\ Brown}
\email{nmbrown@uchicago.edu}
\affiliation{ 
Department of Physics, University of Chicago, Chicago, IL 60637
}%
\affiliation{ 
James Franck Institute, University of Chicago, Chicago, IL 60637
}%

\author{Bryan VanSaders}
\altaffiliation[Now at]{ 
Department of Physics, Drexel University, Philadelphia, PA 19104
}%
Department of Physics, University of Chicago, Chicago, IL 60637
\affiliation{ 
James Franck Institute, University of Chicago, Chicago, IL 60637
}%

\author{Jason M.\ Kronenfeld}
\affiliation{ 
Department of Chemistry, Stanford University, Stanford, CA 94305
}%

\author{Joseph M.\ DeSimone}
\altaffiliation[Also at]{ 
Department of Radiology, Stanford University, Stanford, CA 94305
}%
\affiliation{ 
Department of Chemical Engineering, Stanford University, Stanford, CA 94305
}%

\author{Heinrich M.\ Jaeger}
\affiliation{ 
Department of Physics, University of Chicago, Chicago, IL 60637
}%
\affiliation{ 
James Franck Institute, University of Chicago, Chicago, IL 60637
}%

\date{\today}

\begin{abstract}
Acoustic levitation is frequently used for non-contact manipulation of objects and to study the impact of microgravity on physical and biological processes.
While the force field produced by sound pressure lifts particles against gravity (primary acoustic force), 
multiple levitating objects in the same acoustic cavity interact via forces that arise from scattered sound (secondary acoustic forces).
Current experimental techniques for obtaining these force fields are not well-suited for mapping the primary force field at high spatial resolution and cannot directly measure the secondary scattering force.
Here we introduce a method that can measure both acoustic forces \emph{in situ}, including secondary forces in the near-field limit between arbitrarily shaped, closely spaced objects. 
Operating similarly to an atomic force microscope, the method inserts into the acoustic cavity a suitably shaped probe tip at the end of a long, flexible cantilever and optically detects its deflection. 
This makes it possible to measure forces with a resolution better than 50 nN, and also to apply stress or strain in a controlled manner to manipulate levitated objects.
We demonstrate this by extracting the acoustic potential present in a levitation cavity, directly measuring the acoustic scattering force between two objects, and applying tension to a levitated granular raft of acoustically-bound particles in order to obtain the force-displacement curve for its deformation.

\end{abstract}

\maketitle

\section{Introduction}
Intense ultrasound can be used to levitate small objects, lifting them against the force of gravity by transfer of momentum between the sound field and the object.
This phenomenon, known as acoustic levitation, is used for a wide range of non-contact manipulation tasks.
Fluid droplets loaded with solutes can be mixed and analyzed without a container or substrate by levitation.\cite{Honda_Fujiwara_Hasegawa_Kaneko_Abe_2023, Santesson_Nilsson_2004}
Acoustic levitation can be used to emulate microgravity and its effects on different processes,\cite{Santesson_Nilsson_2004, Vashi_Yadav_Nguyen_Sreejith_2023} such as the development of microstructure in a solidifying metal alloy,\cite{Yan_Hong_Geng_Wei_2015} or to mimic the atmospheric conditions of airborne dust particles.\cite{Méndez_Harper_Harvey_Huang_McGrath_Meer_Burton_2022}
It can also be used in microgravity conditions to trap and manipulate objects.\cite{Wang_Anilkumar_Lee_Lin_1994, Dumy_Jeger-Madiot_Benoit-Gonin_Mallouk_Hoyos_Aider_2020}
In the life sciences, the technique has been used to stably hold protein samples on thin films for crystallographic analysis,\cite{Kepa_Tomizaki_Sato_Ozerov_Sekiguchi_Yasuda_Aoyama_Skopintsev_Standfuss_Cheng_2022}
grow layered structures of neural cells,\cite{Bouyer_Chen_Güven_Demirtaş_Nieland_Padilla_Demirci_2016}
and manipulate living creatures ranging from bacteria to zebrafish embryos and insects.\cite{Akkoyun_Gucluer_Ozcelik_2021, Xie_Cao_Lü_Hong_Wei_2006}
Understanding the forces experienced by a levitated object is important not only when dealing with delicate samples (e.g.\ living organisms), but also to optimize contactless manipulation of multiple levitated objects.\cite{Abdelaziz_Grier_2020, St_Clair_Davenport_Kim_Kleckner_2023}
Furthermore, in all of these cases it is vital to understand the interactions between levitated objects and the mode structure of the acoustic resonance cavity.\cite{Andrade_Polychronopoulos_Memoli_Marzo_2019, Rudnick_Barmatz_1990}

An object can be levitated in air by creating a sufficiently intense acoustic standing wave between an ultrasound source and a reflector.
For a single particle levitated in an acoustic cavity, the acoustic radiation potential is given by the Gor'kov potential
\begin{equation}\label{eq:gorkov}
    U_{\text{rad}} (\mb{r})
    =
    \frac{4 \pi a^3}{3} \br{\frac{1}{2c_0^2 \rho_0} f_0 \left \langle p(\mb{r}, t)^2 \right \rangle - \frac{3}{4} \rho_0 f_1 \left \langle | \mb{v} (\mb{r}, t)|^2 \right \rangle}
\end{equation}
where $a$ is the particle radius, $\rho_0$ is the density of air, $c_0$ is the sound speed in air, $p(\mb{r}, t)$ and $\mb{v} (\mb{r}, t)$ are the pressure and velocity fields, and the angled brackets indicate a time average over a cycle of the acoustic wave.\cite{Gorkov_1962, Lim_VanSaders_Jaeger_2024}
The scattering coefficients $f_0,f_1$ are defined as
\begin{equation}\label{eq:scatteringcoeff}
    f_0
    =
    1 - \frac{c_0^2 \rho_0}{c_p^2 \rho_p}
    ,~~
    f_1
    =
    \frac{2 (\rho_p / \rho_0 - 1)}{2 (\rho_p / \rho_0) + 1}
\end{equation}
where $\rho_p$ and $c_p$ are the density and sound speed for the material of the particle.
The acoustic contrast factor $\Phi = f_0 + \frac{3}{2} f_1$ quantifies how strongly acoustic forces couple to a levitated object and is determined by the densities and isentropic compressibilities of the object and surrounding medium.\cite{Bruus_2012, Wu_VanSaders_Lim_Jaeger_2023}
For any solid or liquid in air, $f_0 \approx f_1 \approx 1$, so $\Phi$ is positive and close to its maximum value of $\frac{5}{2}$. 
As a result the objects are strongly drawn to pressure nodes in the standing wave (unlike, e.g.~air bubbles in a liquid medium, which have $\Phi<0$). 
For 40 kHz ultrasound in air at room temperature, pressure nodes are spaced at half wavelength intervals of $\lambda/2\approx 4.25$ mm.
Confinement to a pressure node plane against the force of gravity is due to what is referred to as the primary acoustic force: the force due to scattering between the cavity mode and the boundaries of the object.
This force is computed from the radiation potential as
\begin{equation}\label{eq:radforce}
    \mb{F}_{\text{rad}}(\mb{r})
    =
    - \nabla U_{\text{rad}}(\mb{r}),
\end{equation}
and as such its exact form depends on the structure of the acoustic field.

A levitated object can also experience forces from sound scattered off nearby objects. 
These  are referred to as second-scattering forces, or simply secondary acoustic forces.
If multiple solid objects are present in the cavity, these secondary acoustic forces can cause short-range interparticle attraction and aggregation.
For two solid spheres that are much smaller than the acoustic wavelength and confined to the levitation plane, the secondary interparticle force is attractive and at close approach scales as
\begin{equation}
    \mb{F}_{\text{sc}}(r)
    \simeq
    - \frac{E_0 a^6}{r^4} \hat{r}, 
    \label{eq:silvabruus}
\end{equation}
where $r$ is the center-to-center distance and $E_0$ represents the acoustic energy density.\cite{Silva_Bruus_2014, Lim_VanSaders_Jaeger_2024}
Note that the magnitude of the secondary scattering force is proportional to the product of the two particles' volumes.
While attractive in the nodal plane, the interaction is repulsive normal to that plane, i.e.\ along the levitation axis.
When many objects are levitated, this results in the formation of a close-packed, quasi-two-dimensional membrane or raft.\cite{Lim_VanSaders_Souslov_Jaeger_2022}
For more detail on the three-dimensional structure of the secondary acoustic scattering potential, we refer to Lim \textit{et al.}\cite{Lim_VanSaders_Jaeger_2024}

While a rich theoretical understanding of primary and secondary acoustic interactions involving point scatterers has been developed, there are many commonly-encountered situations without complete theories.
An ideal vertical standing wave generates a nodal plane of infinite horizontal extent.
However, the finite size of an experimental apparatus creates a finite-range acoustic field subject to specific boundary conditions (i.e.~open or closed cavity walls).
This leads to a horizontal gradient of the acoustic potential and, in turn, generates primary acoustic forces parallel to the nodal plane, which act to confine levitated bodies to the center of the cavity.\cite{Andrade2010}
Furthermore, objects which occupy a non-negligible fraction of the cavity volume may cause changes to the cavity modes that are dependent on the position and velocity of levitated bodies.\cite{Rudnick_Barmatz_1990, Andrade_Polychronopoulos_Memoli_Marzo_2019}
Due to these complications, the acoustic fields within realistic cavities are typically understood with numerical simulation.

Secondary interactions are well-approximated by scattering theory for pairs of objects much smaller than the acoustic wavelength (so that object geometry can be neglected), or that are highly symmetric (e.g.~spheres).
It is substantially more difficult to derive analytic expressions for finite-sized anisotropic objects or systems of many particles.
For example, two levitated cubes will align such that they make contact along their edges, because the acoustic interactions are enhanced in the vicinity of the sharp edge.\cite{Lim_Murphy_Jaeger_2019}
For objects whose size approaches the scale of the viscous boundary layer ($\approx$ 11 $\upmu$m for 40 kHz ultrasound in air), the viscosity of the fluid starts to play a  role and results in acoustic microstreaming.
This introduces an additional, repulsive interparticle force.
Following from the balance of attractive scattering and repulsive microstreaming forces, such particles are separated by a fixed distance.\cite{Wu_VanSaders_Lim_Jaeger_2023}
Investigation of these effects also typically requires numerical simulation.
Ideally, simulations and theory would be complemented by experimental measurements of acoustic interaction forces.
However, determining the fine structure of acoustic forces on small length scales is challenging.

As a proxy to directly measuring acoustic forces, the local velocity and pressure fields could be measured.
Commercial ultrasound sensors that can directly measure acoustic pressure are available in sizes as small as 5 mm.
However, this is still too large to avoid significantly altering cavity resonance and mode structure.
Other methods of experimentally mapping the fields within an acoustic cavity include schlieren imaging \cite{Contreras_Marzo_2021} and thermography.\cite{Onishi_Kamigaki_Suzuki_Morisaki_Fujiwara_Makino_Shinoda_2022}
Schlieren optics can resolve local pressure differences in the air, while thermography can detect local temperature changes resulting from differing air velocities in the acoustic field.
These techniques work well for large-volume fields in open space, where multiple nodes are present.
However, these imaging techniques can be difficult to use in cavities with a single node, or where the more interesting field variations are within the nodal plane.
Because both methods are affected by heat transfer in air, the presence of other heat sources can impact the accuracy of these measurements.
Additionally, in all the above techniques, a conversion (dependent on a theoretical model) between pressure/velocity fields and acoustic forces must be carried out as a final step.

Here, we introduce a method capable of directly measuring acoustic forces and mapping out the acoustic potential.
This method uses 3D-printed probe particles that are connected to thin cantilevers and moved under computer control in $xyz$.
Deflection of the cantilevers (measured with high-speed video) enables force measurement  \emph{in situ}.
By carefully designing the probes, we can localize and limit their disruption to the acoustic field within the cavity.
In many ways, this is similar to the operation of an atomic force microscope, albeit here on larger scales.
The probes can be used to  map out the primary acoustic field in the cavity or measure the secondary scattering force between probes and various levitated objects.
Beyond measuring forces, suitably designed probes can also be used to \emph{perturb}  acoustically levitated granular matter and measure its mechanical response. 
An example is shown in Fig.\ 1, where two probes are inserted from different sides into the cavity to anchor a levitated granular raft (see inset).
One of these probes can then be used to apply controlled amounts of stress or strain and the other used to measure the raft's mechanical response to the perturbation.
In the following, we will focus on primary forces and secondary scattering effects for objects of size 100 $\upmu$m - 1 mm, for which we can neglect microstreaming.
However, the basic concept still applies in the presence of microstreaming.

\section{Experimental Setup}
\subsection{Levitation system}
\begin{figure}
\includegraphics[width=3.5in]{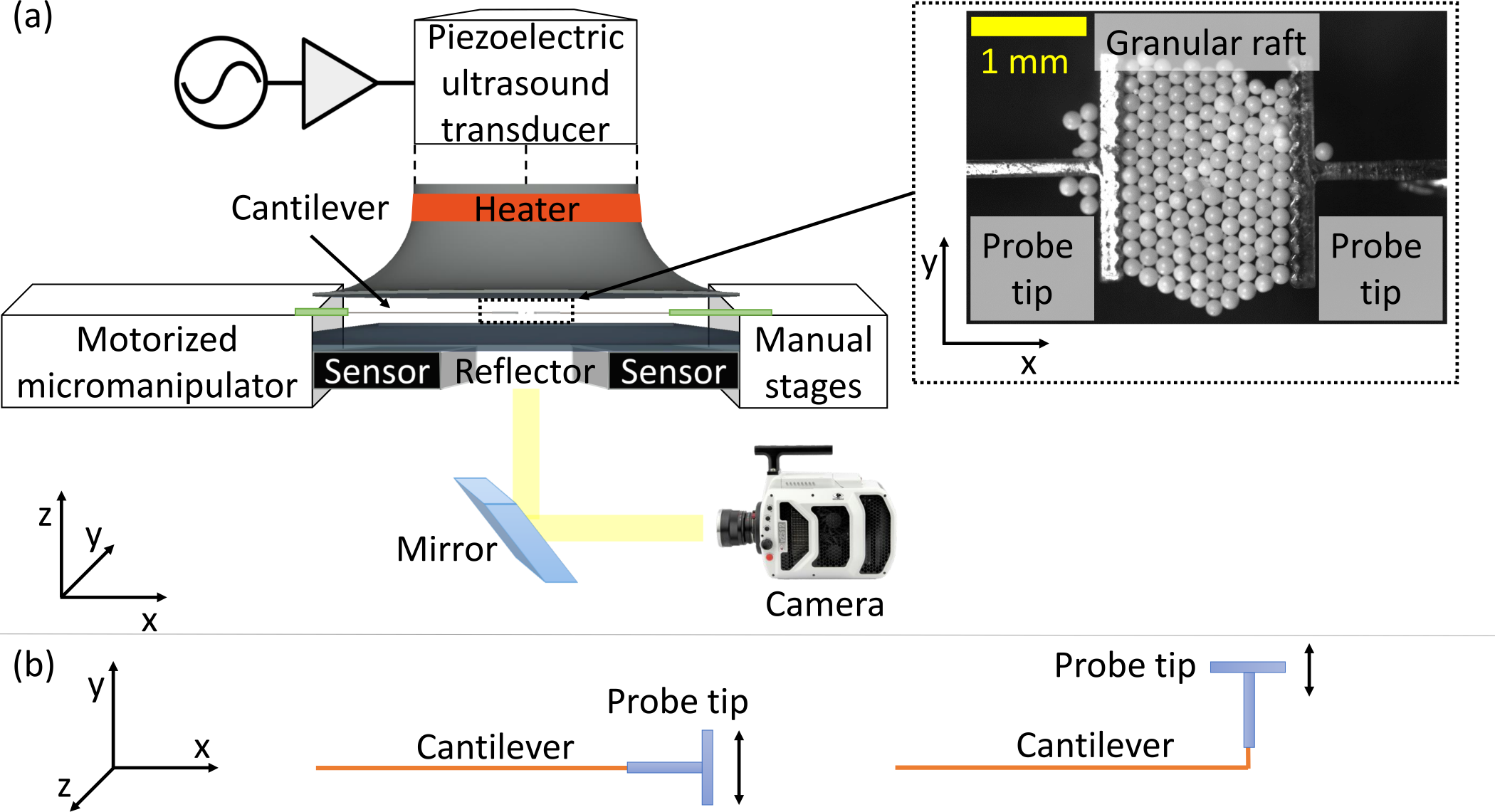}
\caption{Experimental setup. 
(a) Diagram of the experimental system viewed from the front (not to scale). 
A sinusoidal electrical signal is run through a high-voltage amplifier and applied to a piezoelectric ultrasound transducer, which is attached to a machined aluminum horn. 
A standing wave forms in the acoustic cavity between the horn and a transparent reflector plate. 
Four small ultrasound sensors (two are shown) are mounted below the reflector.
Experiments are filmed from below via a mirror with a high-speed camera.
From the left, a motorized micromanipulator can insert and control a force probe on a wire cantilever.
From the right, another probe is inserted using a manual stage. 
Inset: Experimental image of the acoustic cavity, viewed from below.
Levitated granular particles form a raft, which is compressed between two T-shaped probe tips.
The probe tips are attached to thin wires, which act as cantilevers.
After calibrating the cantilevers, forces can be inferred from the positions of the probes.
(b) Schematic of probes in two different orientations  (viewed from above or below, looking onto the levitation plane). 
Both consist of a wire cantilever and a 3D-printed probe tip of any shape.
Probes can measure force in either the $y$- (left) or $x$-direction (right).
}
\label{fig:setup} 
\end{figure}

Figure \ref{fig:setup}(a) shows the experimental setup.
A function generator (BK Precision 4052) outputs a sinusoidal signal which is then amplified by a high-voltage amplifier (A.A. Lab Systems A-301 HS).
This signal drives a piezoelectric ultrasound transducer (Steiner \& Martins SMBLTD45F40H) consisting of two piezoelectric disks between aluminum segments.
A custom-machined aluminum horn mechanically amplifies the acoustic wave (design from Andrade \emph{et al.} \cite{Andrade2010}).
The resonance frequency of the system is $f_0 \approx 40.45$ kHz.
A standing wave forms between the bottom of the horn and a glass slide with a conductive indium tin oxide (ITO) coating.
Both the reflector and the aluminum horn are grounded to minimize electric field contributions.
Typically the spacing between reflector and horn is set to $\lambda/2 \approx 4.25$ mm, thus generating a single pressure node in the $xy$-plane.
The reflector slide sits atop an acrylic box, which contains a silver-coated mirror angled at 45$^\circ$.
This allows us to film levitating structures from below at high frame rate and with a large magnification range using a fast video camera (Phantom V2512) and specialized optics (Navitar Resolv4K lens, working distance $d = 72$ mm).
To hold and move force probes, we use a motorized micromanipulator (Eppendorf Patchman NP2)  together with manual stages.
The setup is surrounded by a $24"\times 12"\times 18"$ acrylic box with aluminum framing and removable doors.
This helps establish  a controlled environment (see below) and suppresses ultrasound propagation into the lab. 
The setup is placed on a passive vibration isolation platform (ThorLabs IsoPlate PTT900600) to minimize mechanical disturbances.

\subsection{Controlling experimental conditions}
The ability to quantify acoustic forces in a levitation cavity reveals that a levitation apparatus as described above can have significant drift in applied acoustic force strength.
A key controlling parameter is temperature, which shifts the resonance conditions as it changes.
As the aluminum horn vibrates, it heats up.
This can slightly change the size and shape of the horn, with pronounced effects on its resonance frequency.
To account for this, the horn is maintained at a constant temperature of 35 $^\circ$C with silicone heater tape and a commercial PID controller (Inkbird ITC-16).

\begin{figure}
    \centering
    \includegraphics[width=3.5in]{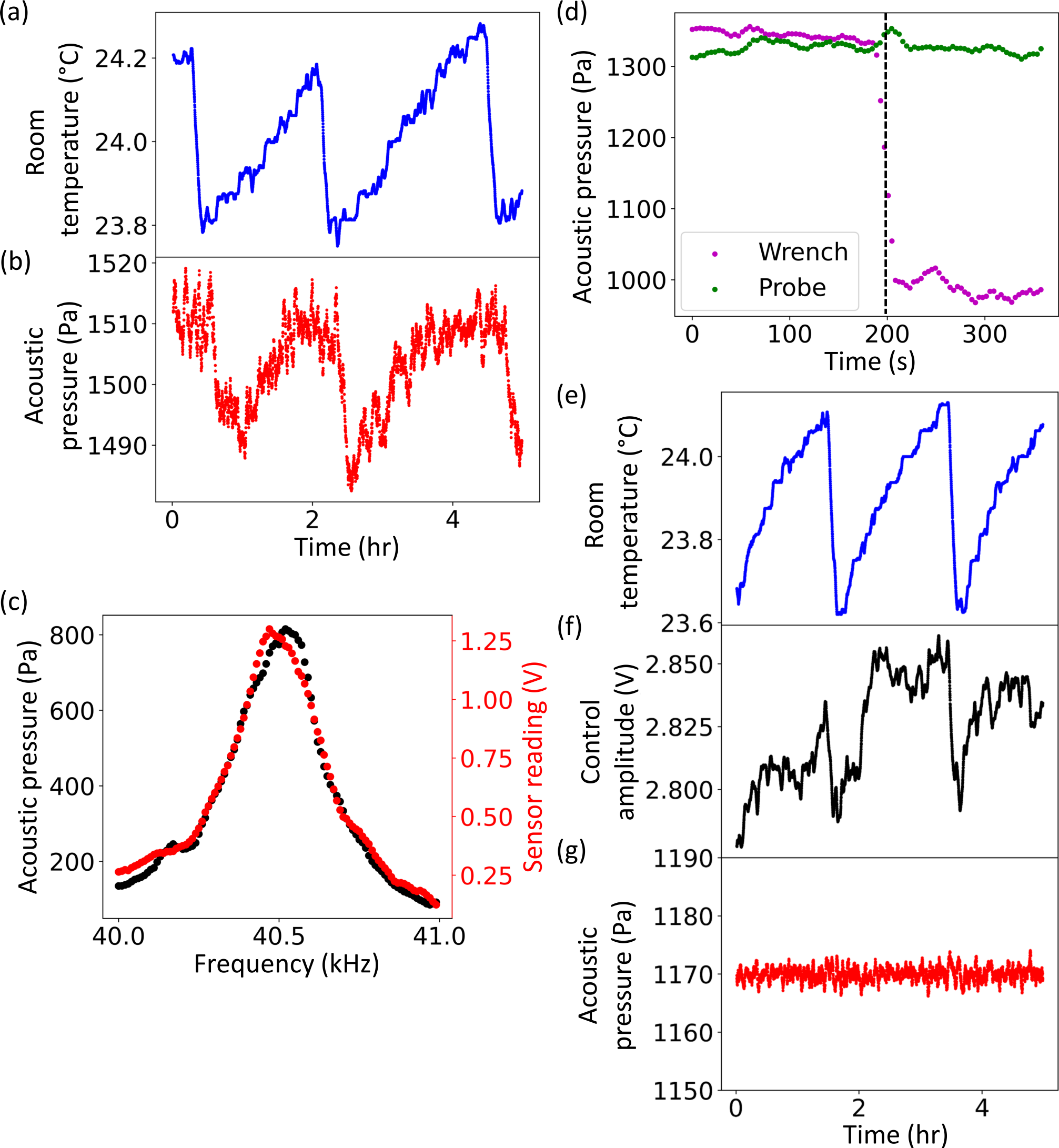}
    \caption{
    Experimental data from acoustic pressure measurements and control.
    (a-b) Fluctuations in room temperature can affect the acoustic pressure in the system. Room temperature (a) changes over the course of hours, resulting in changes to acoustic pressure (b) when uncontrolled.
    (c) Plot of acoustic pressure (black) measured with a calibrated optical microphone placed just outside the acoustic cavity and the voltage (red) measurement from ultrasound sensors placed below the reflector vs.\ frequency of the acoustic wave.
    (d) Plot of acoustic pressure measurements from sensors below the reflector vs.\ time, where a small Allen wrench (magenta) or a force probe (green) is inserted into the cavity (insertion time: vertical dashed line).
    (e-g) Controlling the amplitude of the input AC signal to compensate for environmental changes.
    As the room temperature (e) changes, the control amplitude (f) adjusts to keep the acoustic pressure (g) constant.
    }
    \label{fig:control}
\end{figure}

In addition, the energy density in the acoustic cavity is sensitive to changes in the gap height between the horn and the reflector due to expansion or contraction of the aluminum frame they are both mounted on. 
Changes in room temperature can noticeably affect this gap height, and
while our lab space is climate controlled, throughout a day the temperature can vary as much as 1 $^\circ$C.
Such variation can change the length of a $12"$ aluminum bar by $\approx$ 10 $\upmu$m.
In our experimental setup, even smaller temperature changes (of the order of 0.1 $^\circ$C) can be correlated with variations in the acoustic pressure of up to $10\%$.
This effect can be mitigated by mounting the ultrasound transducer on a structure that is fully enclosed within an insulating box around the setup.
However, the effects of temperature are still noticeable even with these adjustments.
This can be seen in Fig.\ \ref{fig:control}(a, b), which show the room temperature (a) and measured acoustic pressure (b) over a period of hours.
As the room temperature changes by 0.5 $^\circ$C, the acoustic pressure in the cavity changes by 2.7$\%$. 

To further compensate for this, we automatically adjust the acoustic pressure to maintain a constant value throughout an experiment.
Acoustic pressure is measured by four low-profile ultrasound receivers/transmitters (PUI Audio SMUT-1040K-TT) mounted directly below the reflector plate.
Acting as microphones, these sensors output voltage signals proportional to the pressure of the acoustic wave.
We sum the root-mean-square amplitudes of the four sensor signals to obtain a representative value for acoustic pressure.

We use a laser-based pressure sensor (Xarion Eta100 Ultra optical microphone) to characterize the acoustic resonance cavity and examine the frequency response of the PUI Audio ultrasound sensors.
The calibrated optical microphone can measure a broad band of frequencies with high precision, while the ultrasound sensors are optimized for 40 kHz signals and do not have a known calibration.
The optical microphone has a thin Fabry-Perot cavity at its tip which can be inserted into the acoustic cavity; however, most of the microphone head is larger than the typical gap width in the acoustic levitation setup, so the microphone can only be placed into the edge of the cavity instead of closer to the center.
The sensors are $\sim 1000\times$ less expensive than the optical microphone and significantly easier to embed within the apparatus, making them a convenient measurement tool for monitoring overall pressure levels.

We performed a frequency sweep to determine the quality of the resonance cavity and to investigate our two ultrasound measurement devices.
This is plotted in Fig.\ \ref{fig:control}(c).
The optical microphone data (black) and the sensor data (red) show very similar profiles across a range of frequencies.
This suggests that the ultrasound sensor response is roughly constant in this frequency range.    
Matching the input signal amplitude to that used in levitation experiments (below),  we compute a resonance cavity quality factor of $Q \approx 85$.

Given $Q$, we can estimate the sound pressure inside the acoustic cavity formed in the gap between horn and reflector.  To do this, we use that $Q$ can be expressed as
\begin{equation}
    Q
    =
    2 \pi f_0 \frac{E_0}{\dot{E_0}},
    \label{eq:Qfactor}
\end{equation}
where $E_0V$ is the acoustic energy inside a cavity of volume $V$ and $\dot{E_0}V$ is the rate at which this energy is dissipated (note that energy is simultaneously being added to the cavity by the horn's oscillation).
Exciting the cavity with a plane pressure wave and assuming a spatially constant energy density inside, we can write
\begin{equation}
    E_0
    \approx 
    \frac{p_0^2}{\rho c^2},
    \label{eq:cavityenergystored}
\end{equation}
where $p_0$ is the amplitude of the pressure wave, $\rho$ is the air density, $c$ is the sound speed.
In the steady state, the energy dissipation rate $\dot{E_0}V$ is equal to the acoustic power supplied by the horn, which leads to
\begin{equation}
    \dot{E}_0 V 
    =
    \frac{p^2}{\rho c} A
    \label{eq:cavityinputpower}
\end{equation} 
where $p$ is the pressure amplitude produced by the horn when radiating into free space and $A$ is the horn’s cross-sectional area.
From this we obtain an expression that relates the steady-state sound pressure $p_0$ built up in a cavity with quality factor $Q$ to the horn’s free space radiation pressure $p$: 
\begin{equation}
    p_0
    \approx
    \sqrt{\frac{Q c}{2 \pi f_0 h}} p.
\end{equation}
Here we took the cavity volume as $V=Ah$, with $h$ as the gap height between horn and reflector.

Evaluating this with values for $Q$ and $f_0$ from our experimental setup, we compute a pressure ratio  $\frac{p_0}{p} \approx 5.2$.
We can obtain $p$ by using the optical microphone to measure the acoustic pressure directly below the horn when there is no reflector.
For the same input voltage as for the data  shown in Fig.\ \ref{fig:control}(c), this yields $p$ = 450 Pa near the center of the horn.
Thus, the maximum cavity pressure for these data was  $p_0 \approx 2340$ Pa, equivalent to 161 dB.
We can correlate this with our ultrasound sensor readings in Fig.\ \ref{fig:control}(c) to obtain a pressure-voltage conversion factor $\alpha \approx 1800$ Pa/V.
This is the conversion factor used for Fig.\ \ref{fig:control} (a-b, d-g).
Note that this is only an estimate, as the energy density inside a cavity with open sides is not uniform and is instead somewhat larger in the center (see Section \ref{sec:cavity}).

The presence of an object within the system can perturb the resonance of the acoustic cavity.
To demonstrate this, we measure the acoustic pressure while different objects are inserted to and removed from the trap.
Figure \ref{fig:control}(d) shows data from the ultrasound sensors throughout this process.
We inserted a small Allen wrench (diameter 1.8 mm) into the center of the acoustic levitation setup.
These data are shown in magenta in (d): note that the measured acoustic pressure drops by nearly 25\% when the object is present (after the dashed line).
The object significantly disrupts the resonance of the acoustic cavity.
Separately, we inserted a small, custom probe like that shown in Fig.\ \ref{fig:setup}(a, inset) into the otherwise empty cavity.
These data are shown in green.
The acoustic pressure stays nearly the same regardless of the presence of the probe.
Thus, any tools we intentionally introduce into the cavity must be carefully designed for minimal perturbation to the system.

Within the acoustic pressure control system, the signals from the ultrasound sensors are collected by a DAQ (National Instruments USB-6009) and processed on a computer.
The amplitude of the acoustic signal is calculated as the sum of the root-mean-square values of the four sensor signals.
The DAQ's sampling rate is significantly lower than $f_0$ (with four channels in use, the sampling rate is 12 kHz), so data are collected over many cycles ($>1$ s) to avoid the effects of aliasing.
PID control, implemented in Python, is used to adjust the amplitude of the AC signal from the function generator and control the acoustic pressure in the system.
Room temperature, measured near the experimental setup, is incorporated into the PID loop as a feedforward variable.
This helps to maintain a consistent acoustic pressure level both throughout a single experiment and for experiments performed at different times.

Figure \ref{fig:control}(e-g) show data from the control system over a five hour period.
Room temperature (e) fluctuates over a range of approximately 0.5 $^\circ$C during this time.
If we did not compensate for this, it could significantly change the acoustic pressure in the cavity, which is proportional to the square root of the acoustic energy density.
The output control amplitude (f) changes to adjust for the effects of the temperature fluctuation.
The acoustic pressure (g) is maintained at the setpoint despite the changes in environmental conditions, in contrast with the uncontrolled system in (b).

\subsection{Force detection, probe tips, and calibration}
\begin{figure}
    \centering
    \includegraphics[width=3.5in]{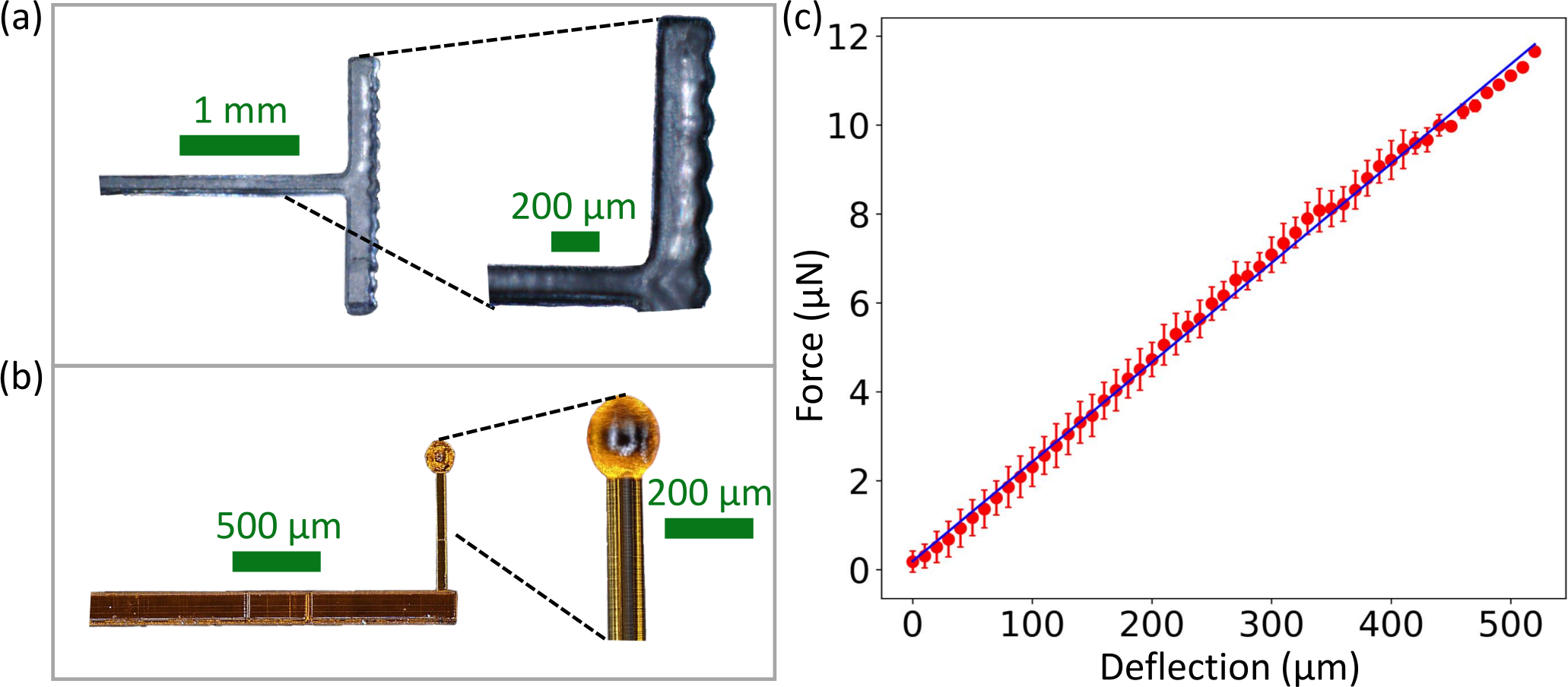}
    \caption{
    Force probes and calibration. 
    (a) 3D-printed probe tip used in tension experiments. The probe is T-shaped so it can be easily attached to a cantilever. The probe edge has particle-size bumps to promote short-range attraction and allow for control over crystal alignment of a granular raft.
    Left image shows the full probe tip, right shows details of the edge.
    (b) 3D-printed probe tip used to map out the secondary acoustic force field due to scattering. 
    It has a 200 $\upmu$m-thick handle for attachment to a cantilever, a 70 $\upmu$m-thick arm, and a sphere on the end of the thin arm that resembles a single granular particle used in the experiments.
    Left image shows the full probe tip, right shows details of the ``particle''.
    (c) Exemplary plot of force vs.\ deflection for  calibration of a force probe. The line (blue) is a fit to the data (red). Error bars indicate one standard deviation.
    }
    \label{fig:forcesensors}
\end{figure}
We measure forces within the acoustic cavity by attaching probe tips to thin copper wire (36 AWG, diameter 127 $\upmu$m) that acts as a soft, flexible cantilever and whose deflection can be tracked optically. 
Different probe tip shapes can be designed for different measurements and perturbations.
For detecting forces experienced by individual levitating particles, we use probe tips shaped similarly to those particles.
Alternatively, for applying forces to whole rafts of levitating objects, we use larger, T-shaped probes.
Examples of these are shown in Fig.\ \ref{fig:forcesensors}(a, b), with images taken with an Olympus DSX1000 optical microscope.

The working surface of the T-shaped probes which contacts a levitated raft is corrugated with the same length scale as the particles which constitute the raft.
These regions of curvature on the probe surface increase the short-range acoustic scattering interaction in a highly localized way.\cite{MelodyPacman} 
By matching the size and spacing of the corrugations with the 180-200 $\upmu$m diameter of the spheres comprising the raft, the spheres are pulled into the concave regions, which enhances the acoustic binding and encourages the formation of a single crystal between probes while also greatly increasing the tangential forces needed to slip the raft relative to the probe.

For measuring the acoustic scattering force in a smaller volume, we use a probe tip with an emulated particle (180 $\upmu$m sphere) on a narrow arm (70 $\upmu$m wide).
A probe tip of this shape is pictured in Fig.\ \ref{fig:forcesensors}(b).
Similarly-shaped probes with larger spheres, to enhance the acoustic force, can be used to map out the acoustic field in the cavity.
In these examples, the bulk of the probe tip geometry, as well as the cantilever wire, lies outside the nodal plane to minimize the impact of the probe body on measurements of the in-plane force (see Section \ref{sec:appsims} for simulations of the impact of the probe arm).

All probe tips are 3D-printed via high-resolution Continuous Liquid Interface Production (CLIP).\cite{Hsiao_Lee_Samuelsen_Lipkowitz_Kronenfeld_Ilyn_Shih_Dulay_Tate_Shaqfeh_etal._2022, Lee_Hsiao_Lipkowitz_Samuelsen_Tate_DeSimone_2022, Kronenfeld_Rother_Saccone_Dulay_DeSimone_2024}
CLIP, a vat photopolymerization-based additive manufacturing technique, employs a digital micromirror device to pattern ultraviolet light (UV; 385 nm) into a sequence of 2D images (2560 $\times$ 1600 pixels, simultaneously) which describe a 3D structure.
An optically transparent, oxygen-permeable, Teflon window at the bottom of the resin vat facilitates a polymerization-inhibited `dead zone', negating part lamination to the window.
Interleaved movement of a high-precision vertical stage facilitates fabrication of free-form 3D structures.
The printing process, controlled via a custom C++ script, achieves horizontal resolution of either 30 or 2 µm (Digital Light Innovations 3DLP9000 or In-Vision Technologies AG 9000 Firebird Light Engine, respectively).
Vertical fabrication is guided by a high-precision stage (Newport GTS70V or ThorLabs KVS30/M), with layerwise dosage determined by the light penetration depth and critical cure dosage inherent to the resin formulation.
Probes are fabricated from either a TMPTA-based resin (for clear probes) or an HDDA-based resin (for orange probes).
The trimethylolpropane triacrylate (TMPTA) formulation contains 2.5 wt\% diphenyl(2,4,6-trimethylbenzoyl) phosphine oxide photoinitiator and 0.4 wt\% BLS1326 UV absorber, while the 1,6-hexanediol diacrylate (HDDA) formulation contains 0.5 wt\% 1,6-hexanediol dimethacrylate, 5.0 wt\% phenylbis(2,4,6-trimethylbenzoyl) phosphine oxide photoinitiator, and 0.5 wt\% Sudan I UV absorber.

The tips are precisely affixed to the cantilever copper wire using a quick-drying adhesive.
Each cantilever is similarly bonded to a plastic hook or handle that is then secured to a micromanipulator, allowing for precise movement control.
These cantilevers are inserted into the acoustic field and maneuvered with computer or joystick control via a micromanipulator or manually with a micrometer stage.
The position of a probe tip can then be tracked in high-resolution, high-speed video to measure deflection and directly observe the effect of acoustic forces on the probe when inserted into the acoustic cavity.

A convenient way to calibrate the forces experienced by the probes  employs a $\upmu$g-resolution scale (Sartorius CPA26P).
During calibration, the plastic handle at the far end of the cantilever wire is attached to a micrometer oriented along the vertical, $z$-direction and lowered so that it barely makes contact with the scale's platform.
The scale's doors are closed as much as possible, and openings covered with plastic sheeting, to minimize noise from air currents in the room.
Incrementally pressing the probe tip onto the scale, mass readings from the scale are recorded and converted to force.
The force-deflection data are linear to excellent approximation and fit to a line.
The resulting slope is the constant that relates maximum cantilever deflection (or deflection at the probe tip) to measured force.
A sample calibration curve is plotted in Fig.\ \ref{fig:forcesensors}(c).
Alternatively, these probes could be calibrated dynamically by determining the resonance frequency of the cantilevers; we estimate that these sensors have a resonance frequency in the range of 1-10 Hz.
The static loading calibration method, however, has the advantage of being independent of any particular model or relationship between applied force and probe deflection.

The maximum force resolution achievable with this method depends on video resolution, probe materials, and cantilever geometry.
Higher video resolution allows smaller deflections to be observed.
The best video resolution that can be achieved with our current optical setup is 2 $\upmu$m per pixel. 
A system built to use a lens with a shorter working distance could potentially achieve better video resolution, with the trade-off of reduced ability to track vertical movements of objects of interest due to a shallower depth of field.
Additionally, a typical probe tip weighs between 0.1-1.5 mg depending on the tip volume.
These tips then exert a gravitational force of 1-15 $\upmu$N on a cantilever.
This is similar to the magnitude of the forces we measure in experiments in the nodal plane, and can result in vertical deflection of the cantilever by up to a few hundred $\upmu$m. 
Longer cantilevers allow for more sensitivity in the intended measurement direction ($x$ or $y$), but also result in more movement in the $z$-direction; this could affect the reliability of the measurements.
Longer cantilevers also require more settling time to adjust to force changes and are more susceptible to disturbance by any external air currents.
Here, we use sections of copper wire as cantilevers, as the material is readily available and easy to work with.
Cantilevers could be constructed with different cross-sections, or from composite materials, to create sensors with higher sensitivity in some directions than others (i.e.\ small vertical deflections and large in-plane deflections).
This could increase sensor resolution in the nodal plane without resulting in unintended $z$-displacement.
Alternatively, microfabricated MEMS-based sensors with stiff probe arms could be utilized instead of cantilevers, but would necessitate complex and costly fabrication.\cite{Wei_Xu_2015, Sun_Fry_Potasek_Bell_Nelson_2005}
Our methods provide an accessible way to perform $\upmu$N-scale force measurements within an acoustic levitation system.

\subsection{Experimental procedure}
The experimental setup, including the electronics that power the acoustic field and the temperature control system, is allowed to run for at least 30 minutes before any experiments are performed. 
This allows the aluminum horn's temperature to stabilize.
Next, the cavity gap height is adjusted to approximately one-half acoustic wavelength ($\approx4.25$ mm), then fine-tuned to maximize the reading from the ultrasound sensors.
If the experiment involves a granular raft, then particles are inserted into the center of the cavity with a mesh spatula, after which they rapidly self-assemble into a membrane.
Force probes are moved close to the center of the cavity with motorized or manual manipulation.
Because the presence of objects in an acoustic cavity can affect its resonance characteristics,\cite{Andrade_Polychronopoulos_Memoli_Marzo_2019} the gap height is readjusted to maximize ultrasound sensor readout.
Finally, the feedback control for acoustic pressure is switched on and allowed to adjust until it steadily maintains its setpoint.
The probes are moved to perform experiments and recorded with high speed video. 
Pressure and temperature control remain on to ensure consistent experimental conditions.

\section{Mapping an acoustic cavity}\label{sec:cavity}
\begin{figure}
    \centering
    \includegraphics[width=3.5in]{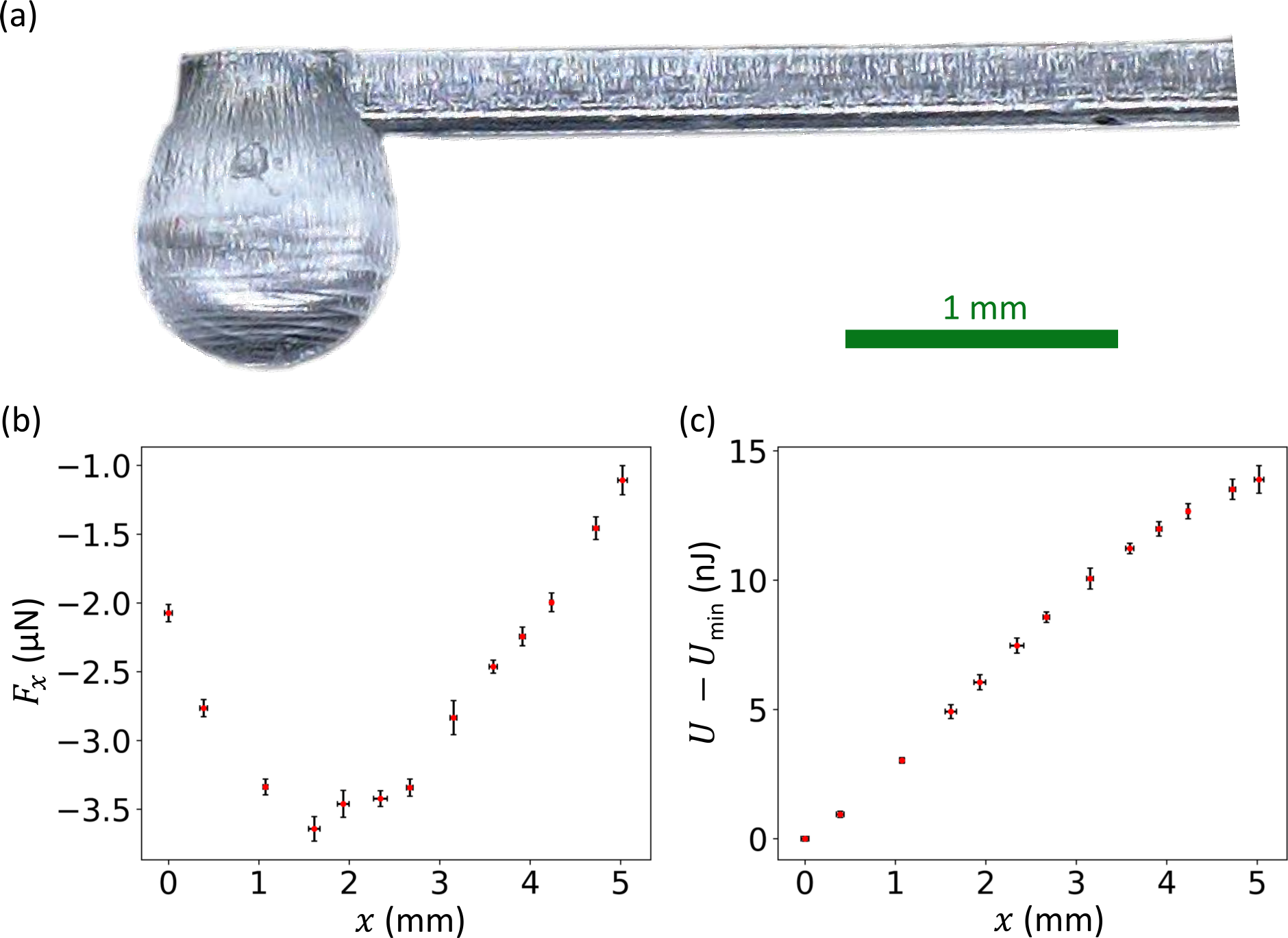}
    \caption{Experimentally mapping the acoustic potential. 
    A spherical probe tip is moved within the acoustic cavity, and its position is recorded with the sound on and off to determine the acoustic force on the probe.
    (a) Image of the probe tip.
    (b) Plot of $F_x$ vs.\ $x$ (error bars: standard error).
    As the probe is moved toward the center of the cavity, the attractive force on it becomes more negative, i.e. stronger, until it reaches a minimum.
    The force then weakens as the probe tip nears the center.
    (c) Acoustic potential $U - U_{\text{min}}$ vs.\ $x$ derived from the data in (b), with well depth $U_{\text{min}}$ (error bars: derived from standard error of $F_x,~x$). 
    As the probe moves towards the center of the cavity, the acoustic potential decreases to a minimum.  
    }
    \label{fig:acousticcavity}
\end{figure}

To characterize the shape of the acoustic potential in the cavity we can spatially map the (primary) acoustic forces on a spherical probe.
To this end we move a probe tip shaped similar to a single particle with a multi-axis micromanipulator to various $(x,y)$ locations in the pressure nodal plane.
The probe position is then recorded with the acoustic field on, and subsequently with the field off.
The difference in the probe positions corresponds to the acoustic force exerted on the probe.
Repeating this process at different positions makes it possible to map out a force field that describes the nodal plane.

The results of a mapping of this kind are shown in Fig.\ \ref{fig:acousticcavity}.
Here a probe shaped like that pictured in Fig.\ \ref{fig:acousticcavity}(a) was used, with a 900 $\upmu$m diam.\ sphere as the tip.
The sphere was positioned with its equator in the nodal plane, keeping the rest of the probe tip out of plane.
Using a larger probe has the advantage of experiencing a larger primary acoustic force (as primary forces scale with the volume of the scatterer, \cite{Wu_VanSaders_Lim_Jaeger_2023} see Eq.\ \ref{eq:gorkov}) and also makes it easier to locate in an image, which results in increased relative resolution for determining force values.
This probe tip was moved from the center of the cavity along the $x$-axis to a distance greater than $\lambda / 2$ and back twice, for four separate trials.
For each measurement, we moved the probe into position and then waited for the acoustic pressure control to adjust to changed resonance conditions before recording observations.
Videos were recorded with the field on and off for each position.
For this and other experiments, image processing to locate probe positions in images was performed with Meta's Segment Anything model.\cite{kirillov2023segany}

Figure \ref{fig:acousticcavity}(b) shows that
as the probe moves towards the center of the cavity, it experiences an  attractive force (indicated by the negative sign) of increasing magnitude until it reaches a minimum of $F_x = -3.64 \pm 0.09$ $\upmu$N at a distance of $x = 1.61 \pm 0.06$ mm from the cavity center.
The magnitude of the force then decreases as the probe gets closer to the center of the cavity.
By symmetry the force vanishes at $x = 0$, though the true cavity center is difficult to locate with an extended object like this particular probe tip.
The exact center of the cavity can be determined best by levitating a small, freely floating object and observing the position of its center of mass.
The force data in Fig.\ \ref{fig:acousticcavity}(b) can be integrated with respect to $x$ to compute the shape of the acoustic potential, plotted in \ref{fig:acousticcavity}(c).
The potential has a minimum at $x=0$, and increases as the probe moves further from the center. 
This increase  tapers off for $x \gtrsim \lambda/2$.

Note the scales of both the force and the potential energy: the in-plane forces experienced by objects in this acoustic trap are on the order of \uN and the energy is in the nJ range.
The video resolution for these experiments was approximately 9 $\upmu$m per pixel.
For the force sensor used, this corresponds to a measurement resolution of 0.19 $\upmu$N.

This characterization of the primary force field in the cavity's nodal plane is vital for planning levitation  experiments in a specific  setup, particularly ones that involve large granular rafts or other extended objects.
For most experiments, we would prefer to stay close to the bottom of the potential well not only to be able to levitate objects reliably, but also to avoid significant variation in the local energy density of the acoustic field.
For this reason, the experiments described in later sections were performed in the region where $x < 2$ mm.
These measurements also reveal the scale of differential primary acoustic forces on extended levitated objects.
Sufficiently soft or large levitated objects can be deformed by these forces. \cite{Dong_Wang_Granick_2019}

\section{Measuring the scattering force}\label{sec:scatt}
\begin{figure}
    \centering
    \includegraphics[width=3.5in]{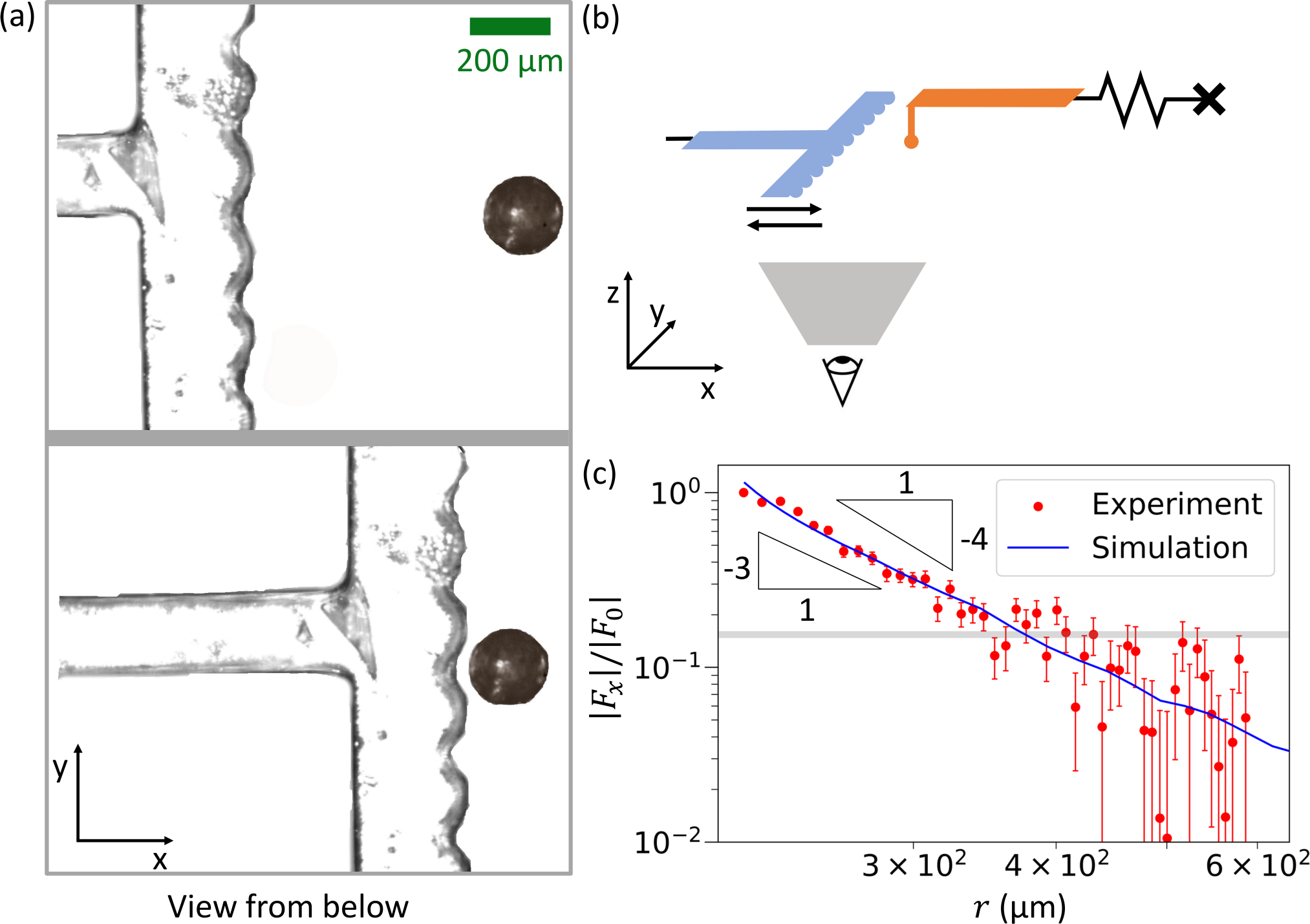}
    \caption{
    Measuring the acoustic scattering force between two objects in the acoustic trap.
    A T-shaped probe is moved towards a force sensor with a 180 $\upmu$m spherical tip and then pulled away. 
    The spherical probe is on a cantilever to measure forces.
    The probe positions are tracked to experimentally determine the acoustic scattering force between the probes.
    (a)
    Images from the experiment, viewed from below.
    (a, top)
    Start of the experiment when the probes are well-separated.
    (a, bottom)
    Middle of the experiment when the two probes are almost in contact.
    Note that the spherical probe has moved slightly to the left as a result of attraction to the larger probe.
    (b) 
    Schematic of the experimental procedure.
    The T-probe moves to approach the spherical probe; the spherical probe is attached to a spring-like cantilever to measure the attractive scattering force between the two objects.
    The eye symbol represents the perspective of the experimental images in panel (a).
    The long arm of the right probe is not visible in panel (a) because it sits outside the focal plane.
    (c)
    Plot of force vs.\ distance for this experiment (red) and corresponding simulations (blue).
    Five trials of the experiment were performed, and the data binned by inter-probe distance (error bars: standard error). 
    The scattering force decreases sharply as distance increases.
    The gray bar marks the noise floor for the experimental force sensor.
    The force is normalized to allow for comparison between experiment and simulation. 
    The shape of the simulation results is similar to the experimental results.
    Two triangles are drawn to represent $r^{-3}$ (left) and $r^{-4}$ (right) power laws; these correspond to a simple analytical model of the scenario.
    }
    \label{fig:scatt}
\end{figure}

We can use probes of different shapes (as in Fig.\ \ref{fig:forcesensors}(a, b)) to directly investigate the acoustic scattering forces between objects of arbitrary shape in the acoustic cavity.
Figure\ \ref{fig:scatt}(a) shows two snapshots from one such experiment, which operates similar to approaching a sample surface with an atomic force microscope tip.
In this example we use as the object to be examined the corrugated T-shaped probe (Fig.\ \ref{fig:forcesensors}(a)) and as the measurement probe a sphere  of  matching size on the end of a thin arm (Fig.\ \ref{fig:forcesensors}(b)).
The particle-like measurement probe is positioned at the center of the acoustic cavity in the nodal plane and connected to a flexible cantilever, while the T-shaped probe, which represents the sample in the corresponding atomic force microscope experiment, is rigidly connected to a motor-driven manipulator and initially positioned further away but also in the nodal plane.
The T-shaped probe is then moved towards the measurement probe at  constant speed until the probes make contact. 
Still images from this process are shown in Fig.\ \ref{fig:scatt}(a), and a schematic is drawn in Fig.\ \ref{fig:scatt}(b).
During the approach, the secondary acoustic force between the two objects can be determined from the position of the particle-like measurement probe (note its slight movement to the left in the lower image).
When the two probes get very close, the measurement probe, being mounted on the flexible cantilever, quickly moves to snap into contact.
The rigidly mounted probe is then moved away until the right probe snaps off and both are again well-separated ($>$ 400 $\upmu$m apart) so that the approach can be repeated.

Five trials of this experiment were performed, and the data binned and averaged by distance between probe centers. 
The results are plotted in Fig.\ \ref{fig:scatt}(c, red), with the error bars representing standard error.
Data corresponding to the probes being in direct contact was removed due to excess noise.
When the probes are nearly in contact, the maximum force measured was $0.227 \pm 0.005$ $\upmu$N.
The experimental images in Fig.\ \ref{fig:scatt}(a) have a resolution of 2.6 $\upmu$m per pixel.
This corresponds to a force resolution of 35 nN per pixel for this particular force sensor (represented by the horizontal gray bar in Fig.\ \ref{fig:scatt}(c)).
The forces are scaled by the maximum force measured in order to match the energy density of the experimental cavity with that of simulations.

As distance increases, the near-field scattering force experienced by the measurement probe decreases quickly.
For two spheres this force decays as $r^{-4}$ (Eq.\ 1). 
However, in this example the T-shaped probe is non-spherical and highly anisotropic, which at close approach leads to a different power law exponent.
Next, we recapitulate this result with finite element method simulations and analytical estimates.

\section{Simulations and Modeling} \label{sec:appsims}

Finite element simulations (COMSOL, Acoustics module) can be used to calculate the acoustic scattering forces on arbitrarily shaped, finite sized objects, as well as the impact of various probe geometries on the acoustic modes present in the cavity.
We simulate an idealized, three-dimensional experimental setup with a flat, vibrating top plate to represent the transducer and horn, and a flat reflector. 
The simulated cavity is a $2" \times 2" \times \frac{\lambda}{2}$ rectangular prism, and the size of scatterers in the field is less than $10\%$ of the cavity width.
The cavity side walls have free slip boundary conditions.
The effect of the scatterers on the acoustic potential is localized to a small volume surrounding the object.
The 3D models of the probe tips were created in AutoCAD, and were based on the models used for 3D-printing the real probe tips used in the experiments.

\emph{Scattering forces.} The power law dependence of the scattering force shown in Fig.\ \ref{fig:scatt} can be recreated in finite element method simulations.
A corrugated T-shaped probe with particle-sized ridges was placed at the center of the simulated cavity, and a particle of corresponding size was placed at various distances from the probe.
The acoustic radiation force on an object can be computed by integrating the momentum flux over its surface.\cite{Settnes_Bruus_2012, Muller_Bruus_2014}
Because the objects are in the nodal plane in a flat acoustic potential, the scattering force has by far the most prominent contribution to the total in-plane force experienced.
Thermoviscous effects can be neglected due to the large size of the objects.
The scattering force on the particle was calculated for each configuration and is plotted in blue in Fig.\ \ref{fig:scatt}(c).
The data from simulation and experiment are in excellent agreement, and both are consistent with a power law decay $r^{-\gamma}$ with exponent $\gamma \approx$ 3.
This agreement shows that the near-field scattering force between an arbitrarily shaped object and a free-floating sphere can be reliably obtained using a particle-shaped probe.
This result demonstrates that the cantilever assembly used to support the particle-shaped probe has a minor impact on near-field acoustic forces.

To further understand the power law decay of the scattering force for this particular experimental configuration, we consider a simple analytical model.
The scattering force between two spheres significantly smaller than the acoustic wavelength is expected to depend on interparticle distance like $r^{-4}$, as in Eq.\ \ref{eq:silvabruus}.
As far as the scattering of sound between the T-shaped probe and the single particle measuring probe is concerned, we model only the top bar of the T and treat it as a line of length $2 \ell$ centered at the origin and oriented along the $y$-axis.
We assume that each infinitesimal piece $dy$ of the T-probe provides a contribution to the scattering force that depends only on its distance from the particle like $r^{-4}$, and that these contributions can be added pairwise to obtain a final result for the force on a particle.
The magnitude of the contribution to the force from an infinitesimal element of the probe is
\begin{equation}
    dF
    \simeq
    \frac{(E_0 a^3 \beta) dy }{r^4}
    =
    \frac{(E_0 a^3 \beta) dy}{(x^2 + y^2)^2} 
\end{equation}
for a constant $\beta$ which depends on the geometry of the probe and the acoustic frequency, directed along the vector connecting the element and the particle.
For a particle positioned at $y=0$, the $y$-component of scattering forces can be neglected by symmetry.
The $x$-component of the differential force is
\begin{equation}
    dF_x
    =
    dF \times \frac{x}{r}
    \simeq
    \frac{(E_0 a^3 \beta) x~dy }{(x^2 + y^2)^{5/2}}. 
\end{equation}
Integrating this expression from $y=- \ell \to \ell$ gives the result
\begin{equation}
    F_x
    \simeq
   (E_0 a^3 \beta)  \frac{2(2 \ell^3 + 3 \ell x^2)}{3 x^3 (x^2 + \ell^2)^{3/2}}.
\end{equation}
Examining the limiting cases, we have
\begin{equation}
    F_x(x \ll \ell) \to x^{-3}
\end{equation}
\begin{equation}
    F_x(x \gg \ell) \to x^{-4}   . 
\end{equation}

Triangles representing these $r^{-4}$ and $r^{-3}$ power laws are drawn in black in Fig.\ \ref{fig:scatt}(c).
Over the range of separations $r$ tracked in these experiments, there is no significant difference between the data from experiments, simulation, and the $r^{-3}$ power law, indicating that the force the particle experiences at this position is dominated by the scattering in its immediate surroundings.

\begin{figure}
    \centering
    \includegraphics[width=3.5in]{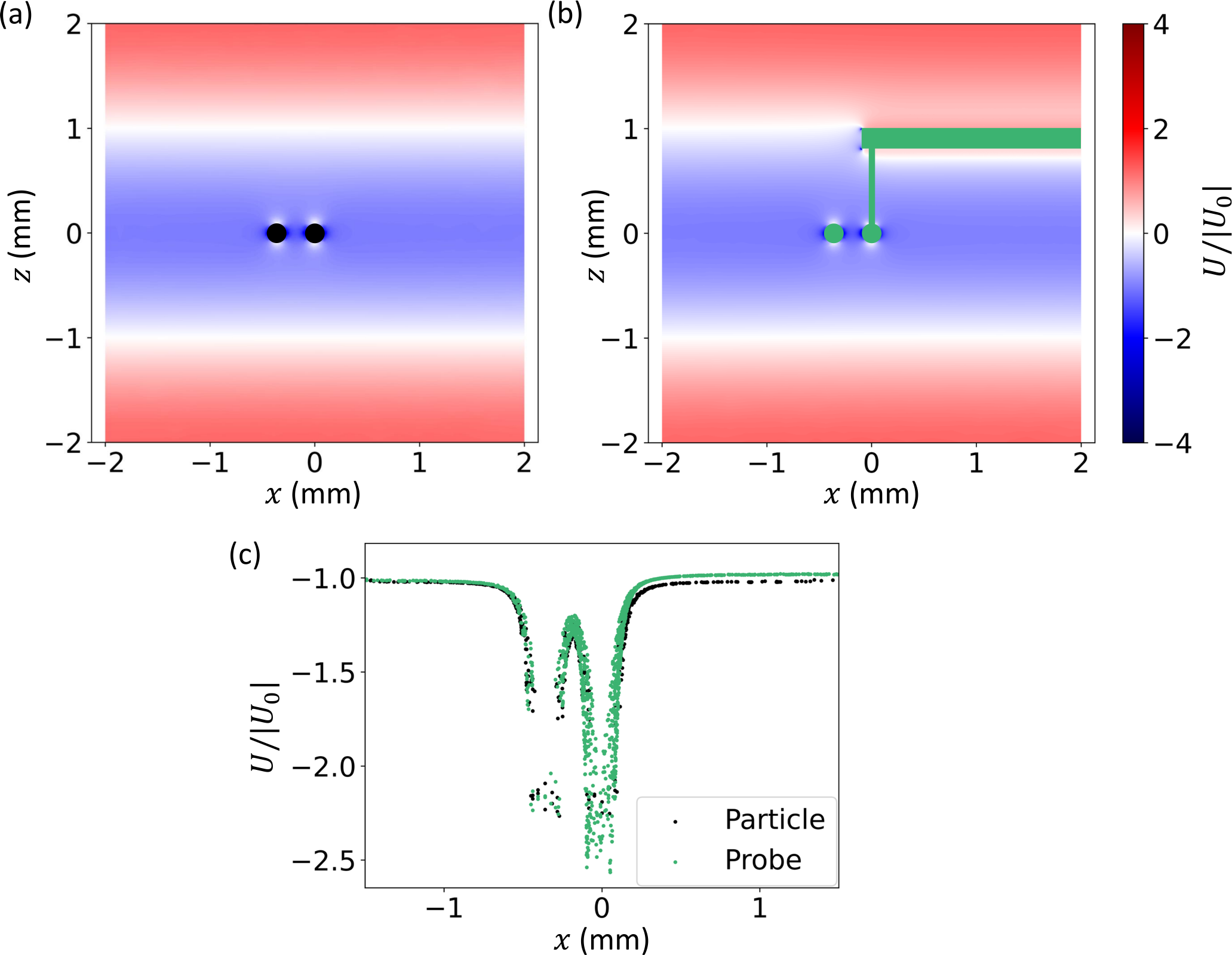}
    \caption{
     Calculations of the acoustic potential energy in a cavity with (a) two 200 $\upmu$m particles and (b) one 200 $\upmu$m particle and one probe matching the one seen in Fig.\ \ref{fig:forcesensors}(b). 
     The potential is plotted in the $xz$-plane on a color scale from blue (negative) to red (positive).
    $U$ is scaled by $U_0$, the mean energy within the nodal plane far from the objects.
    (c) A plot of $\frac{U}{|U_0|}$ vs.\ $x$ along the $x$-axis for the the configuration with two particles (black, (a)) and that with a particle and a probe (green, (b)).
    The two gaps in data represent the locations of the particles and probe.
    }
    \label{fig:COMSOLsmall}
\end{figure}

\emph{Testing the influence of the probe shape}. Similar finite element simulations can be performed to examine the extent to which the geometry of the single-particle measurement probe tip affects the near-field structure of the acoustic potential.
Probe tips as in Fig.\ \ref{fig:forcesensors}(b) necessarily have arms that connect them to a cantilever.
Sound can scatter from these arms, possibly modifying the acoustic forces on levitated objects through secondary forces or by causing modification to the large-scale cavity mode structure.
To quantify these effects, we study scattering between two spherical particles in the near-field limit and compare this, via simulation, with the case where one of the particles has been replaced by a particle-like probe.  
Figure\ \ref{fig:COMSOLsmall} shows the acoustic potential energy $U$ for the two configurations of scatterers in a simulated acoustic cavity.
Figure\ \ref{fig:COMSOLsmall}(a) shows the acoustic potential in the $xz$-plane for two spheres, and (b) shows the potential for a sphere and a probe like that displayed in Fig.\ \ref{fig:forcesensors}(c) (and used for experimental measurements in the previous section). 
Figure \ref{fig:COMSOLsmall}(c) shows the potential along the $x$-axis (in the nodal plane for $-100$ $\upmu$m $ < y < 100$ $\upmu$m) for both configurations. 
Note that objects inside the acoustic cavity can affect its overall resonance (and therefore the average energy density in the system).
To facilitate the comparison of different scattering configurations, the potential is plotted as $\frac{U}{|U_0|}$ where $U_0$ is the average potential in the nodal plane in a region far from the scatterers.
This normalization accounts for any changes to the cavity resonance resulting from the presence of larger objects.
In experiment, feedback control of transducer driving voltage (based on ultrasonic sensor values) is used to stabilize the cavity energy density as objects are added or removed.
The normalized potential for two spheres agrees closely with the normalized potential for a sphere and a probe.
Thus, the presence of 3D-printed probes with these shapes in the acoustic cavity is similar to there being, instead, particles of comparable size (provided that changes in overall energy density of the cavity are accounted for by feedback control).

\section{Microscale mechanical testing of levitated granular materials}\label{sec:tension}

\begin{figure*}[ht!]
    \centering
    \includegraphics[width=7in]{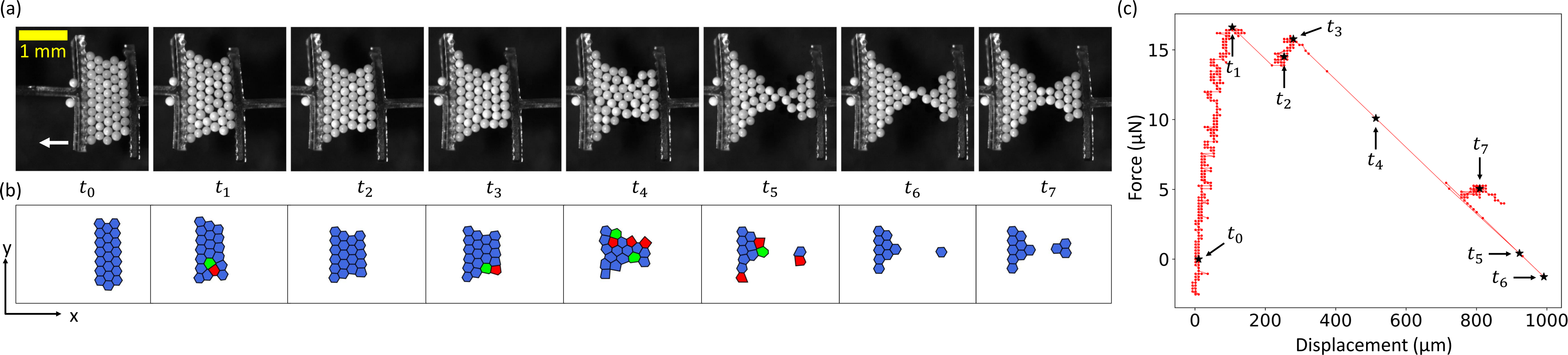}
    \caption{
    Tensile test experiment on a levitated granular raft.
    (a) Time series of experimental images throughout a tensile test.
    The left probe moves in the $-x$ direction at a constant rate  while the right probe measures force via $\pm x$ deflection.
    As the experiment proceeds, the raft deforms, forming a narrow neck.
    (b) Voronoi diagrams of the granular rafts in (a).
    Particles with six nearest neighbors are colored blue, while particles with five and seven nearest neighbors are colored red and green, respectively.
    In 2D, dislocation defects are identifiable as 5-7 defect pairs, one of which can be seen at $t=t_1$.
    (c) Plot of force vs.\ displacement (red) for the same experiment.
    Timestamps $t_0-t_7$ are marked with black stars.
    }
    \label{fig:tension}
\end{figure*}

Using the T-shaped probes pictured in Fig.\ \ref{fig:forcesensors}(a), we can perform experiments that resemble traditional materials testing, albeit here on freely floating structures levitated, self-assembled, and held together by sound-induced forces.\cite{Lim_VanSaders_Jaeger_2024}
As an example, we discuss tensile testing on two-dimensional rafts of levitated granular particles (HDPE, 180-200 $\upmu$m diam.).
The process is shown in Fig.\ \ref{fig:tension}.
We trap an acoustically-bound raft between two T-probes.
The left probe is connected to  a thin, stiff rod, while the right probe is attached to a cantilever which deflects in the $x$-direction to measure tensile force.
The left probe is controlled by a micromanipulator and is first cycled in the $\pm y$-direction multiple times to anneal the sample and align the crystal with the probes.

Once prepared, the left probe is moved in the $-x$-direction at a constant rate of 10 $\upmu$m/s to stretch the raft.
A time series of experimental images of this process is displayed in Fig.\ \ref{fig:tension}(a).
In (b), corresponding Voronoi diagrams are drawn.
The Voronoi cells representing particles with five and seven nearest neighbors are colored red and green, respectively, while particles with six nearest neighbors are colored blue.
Note that particles at the raft edges have been excluded from Voronoi tessellation.
In two-dimensional crystals, a dislocation defect can be identified as a bound 5-7 Voronoi pair.
The left probe applies tensile strain to the raft until it experiences significant plastic deformations, while the right probe measures the tension force experienced by the raft.
At $t_0$, the raft is in a crystalline state.
At $t_1$, it is deforming plastically as indicated by the presence of a dislocation within the sample.
By $t_2$, that slip event has concluded and the raft has extended.
At $t_3$, a second slip event begins.
At $t_4$, $t_5$, and $t_6$, the raft is deforming during a larger slip event that almost severs the raft.
By $t_7$, the raft's two portions have come back together slightly, maintaining a narrow neck and a configuration that is significantly lengthened from the initial state.

Tension force and displacement data for this experiment are shown in Fig.\ \ref{fig:tension}(c) and clearly exhibit the stick-slip behavior observed in macroscopic granular materials\cite{Krim_Yu_Behringer_2011} as well as in nanoscale single-crystal metal pillars.\cite{Greer_Nix_2006}
The zero point of the force is determined by recording the positions of the probes in the acoustic field without the raft (for that, the field is turned off after the conclusion of the experiment so that particles fall, and then turned back on and an image is recorded).
From the start of the experiment ($t_0$), applied force and displacement increase nearly linearly to a peak (near $t_1$).
This corresponds to the cohesive granular crystal deforming via slight internal rearrangements but without particles slipping past one another.
The rapid displacement increase together with a sudden drop in force  after the first peak represents a slip event, and thus the onset of significant plastic deformation.
A local force minimum occurs after the slip as the crystal recovers.
The subsequent rise toward the second peak ($t_2$-$t_3$) signals the buildup in stress before another slip event ($t_4$-$t_6$), during which the raft lengthens significantly and is pulled almost completely apart.
Attraction between the particles (and between the two large, crystalline portions of the raft) causes the raft to remain contiguous  after the event ($t_7$).
The raft contraction between $t_6$ and $t_7$ highlights how the acoustic forces can mimic cohesion due to attractions that are longer-ranged than forces based solely on direct contact.

Experiments like this make it possible to observe and track individual particle displacements that result in global deformation.
We here focused on tensile loading, but in principle the same manipulator setup can be applied to perform tests in other modalities  analogous to traditional mechanical testing, all of which can give insight into the mechanical properties of cohesive granular materials and the interactions between particles in many-body acoustic systems.

\section{Conclusions}

We  developed an experimental approach that can measure forces within acoustic levitation systems with only minimal disruption to the acoustic field by using specially designed microscale force probes (Fig.\ \ref{fig:forcesensors}).
When used in a carefully controlled acoustic system (Fig.\ \ref{fig:control}), this methodology can be used to characterize the acoustic cavity in any given experimental setup (Fig.\ \ref{fig:acousticcavity}).
This provides vital information for planning levitation experiments and can help guide future experimental system design.
It enables direct, \emph{in situ} measurement of the secondary acoustic scattering forces between two objects of arbitrary shape, a scenario that is typically difficult to describe analytically. 
In our current setup, the measurements can be performed with an accuracy better than 50 nN (Fig.\ \ref{fig:scatt}).
Suitably shaped probe tips on the force sensors are found to have a minimal effect on the acoustic cavity structure as demonstrated by simulation (Fig.\ \ref{fig:COMSOLsmall}).
These same tools  not only can probe but also perturb structures assembled by acoustic levitation in a gravity-free, container-less environment. In this way they can be used in experiments that resemble conventional mechanical testing (Fig.\ \ref{fig:tension}).

Here, we have utilized specialized equipment such as a high speed camera, an optical microphone, and advanced 3D printing techniques.
Some of these methods, however, can still be effective without this equipment.
For example, quasi-static measurements could be made using cameras with a slower frame rate.
Simpler probe structures could be constructed by carefully folding thin wires, though high-resolution 3D printing is required for producing probes of the shapes shown in Figs.\ \ref{fig:forcesensors} and \ref{fig:acousticcavity}.

Extensions of this method could use dynamic excitations near the cantilever's resonance frequency to measure acoustic forces, similar to frequency modulation atomic force microscopy techniques.
Additionally, while the experiments described here were performed in air, many biological applications of acoustic levitation involve levitating objects in a liquid or gel.
The acoustic frequencies used in these systems are typically in the MHz range, with wavelengths 10-100$\times$ shorter than the wavelength in our air-based system. 
For use in liquids (e.g.~in microfluidic systems) the tools described here could be miniaturized and supplemented with more sensitive force sensors based on piezoelectrics, and the procedures could be modified to perform similar tests in different acoustic systems.

\begin{acknowledgments}
We thank Brady Wu, Qinghao Mao, and Tali Khain for useful discussions.
This research was supported by the National Science Foundation through award DMR-2104733 and, in part, by the Army Research Office through award W911NF-22-2-0109 (NMB).
We utilized shared equipment and resources at the University of Chicago MRSEC, which is funded by the National Science Foundation under award DMR-2011854. 
This research also benefited from computational resources and services provided by the University of Chicago’s Research Computing Center.
JMK acknowledges support through the National Science Foundation Graduate Research Fellowship Program under grant DGE-1656518. 
\end{acknowledgments}

\section*{Author Declarations}
\subsection*{Conflict of Interest}
The authors have no conflicts to disclose.

\subsection*{Author Contributions}
\noindent\textbf{Nina M.\ Brown}: 
Conceptualization (equal),
Formal analysis (lead),
Investigation (lead),
Methodology (equal),
Software (lead),
Validation (lead),
Visualization (lead),
Writing - original draft preparation (lead),
Writing - review and editing (equal)
\\
\textbf{Bryan VanSaders}:
Conceptualization (equal),
Formal analysis (supporting),
Methodology (equal),
Writing - review and editing (equal)
\\
\textbf{Jason M.\ Kronenfeld}:
Investigation (supporting),
Methodology (supporting),
Resources (equal),
Writing - original draft preparation (supporting)
\\
\textbf{Joseph M.\ DeSimone}:
Funding acquisition (supporting),
Resources (equal),
Supervision (equal)
\\
\textbf{Heinrich M.\ Jaeger}:
Conceptualization (equal),
Methodology (equal),
Formal analysis (supporting),
Funding acquisition (lead),
Resources (equal),
Supervision (equal),
Writing - review and editing (equal)

\section*{Data Availability}
The data that support the findings of this study are available from the corresponding author upon reasonable request.

\appendix


\bibliography{RevisedManuscript_July2024}

\end{document}